\documentclass[superscriptaddress,aps,amsmath,amssymb,showpacs,showkeys]{revtex4-2}
\usepackage{natbib} 
\usepackage[dvips]{graphicx}

\usepackage{times}
\usepackage{braket}
\usepackage{xcolor}
\usepackage{orcidlink}
\usepackage{hyperref}
\usepackage{booktabs}
\usepackage{subfigure}
\usepackage{siunitx}
\hypersetup{
	colorlinks=true,
	urlcolor=magenta,
	linkcolor=red,
	citecolor=blue
}

\begin{document}
\title{A systematic comparison of green valley selection criteria across 
multiparameter spaces using a homogeneous ultraviolet-optical dataset}
\author{Pius Privatus\orcidlink{0000-0002-6981-717X}}
\email[Email: ]{privatuspius08@gmail.com}
\affiliation{Department of Physics, Dibrugarh University, Dibrugarh 786004, Assam, India}
\affiliation{Department of Natural Sciences, Mbeya University of Science and Technology, Iyunga 53119, Mbeya, Tanzania}
\author{Umananda Dev Goswami\orcidlink{0000-0003-0012-7549}}
\email[Email: ]{umananda@dibru.ac.in}
\affiliation{Department of Physics, Dibrugarh University, Dibrugarh 786004, Assam, India}
	
\begin{abstract}
We present a systematic comparison of commonly adopted green valley (GV)
selection criteria by examining their distributions across multiple 
observational and physical parameter spaces. Using a homogeneous 
ultraviolet-optical dataset constructed from the Galaxy Evolution Explorer 
(GALEX) and the Sloan Digital Sky Survey (SDSS), we construct GV samples 
based on rest-frame $u-r$ and NUV$-r$ colours, specific star formation rate, 
and the $D_n(4000)$ spectral index. These samples are analysed in 
colour--stellar mass, colour--magnitude, and star formation rate--stellar mass 
diagrams. We find that the different selection criteria identify statistically 
distinct subsets of GV galaxies occupying different regions of parameter space. 
Ultraviolet-based selections are compact in NUV$-r$ colour space but shift 
toward optically red galaxies and lower star formation activity in the star 
formation rate--stellar mass plane. The $u-r$-selected sample is more tightly 
confined in optical colour space but is biased toward higher star formation 
rates, whereas the $D_n(4000)$-based selection yields the most heterogeneous 
population. In contrast, the sSFR-selected GV sample exhibits the most 
consistent behaviour across all parameter spaces. Despite these differences, 
all selection methods span a similar stellar mass range, indicating that the 
observed variations arise primarily from differences in star formation 
activity rather than stellar mass. The relatively small overlap between the 
different selection criteria demonstrates that GV identification is strongly 
diagnostic-dependent and that the commonly adopted one-dimensional 
definitions are not interchangeable. These results highlight the importance 
of combining complementary diagnostics to obtain a more complete and 
physically meaningful picture of transitional galaxy populations.
\end{abstract}
	

\maketitle    
	
\section{Introduction} \label{secI}
Understanding how and why galaxies stop forming stars and make the transition 
from active systems to quiescent ones remains one of the key 
challenges in galaxy evolution \cite{naab2017theoretical,forster2020star,
fabian2012observational,french2021evolution,saintonge2022cold,
peng2010massquench,kalinova2021star}. The studies carried out over the past 
two decades, including by the Sloan Digital Sky Survey (SDSS) 
\cite{york2000sdss} and the Galaxy Evolution Explorer (GALEX) 
\cite{martin2005galaxy}, have revealed a clear bimodality in galaxy 
populations where the star-forming galaxies occupy the blue cloud, whereas 
quiescent galaxies occupy the red sequence in diagnostic diagrams 
involving the colour–magnitude and colour–stellar mass spaces 
\cite{wyder2007galex,schawinski2014green,strateva2001color,
li2024characterising,baldry2004bimodal,privatus2025main}. 
Between the blue cloud and the red sequence, an intermediate population, 
known as the green valley (GV), lies between the two populations. This region 
is populated by galaxies in a transitional stage, with declining star 
formation as they evolve from the blue cloud to the red sequence. The GV 
provided a useful window for the study of the physical processes responsible 
for the quenching of star formation \cite{brammer2009dead,mendez2011aegis,
levis2025galaxy,ascasibar2025euclid,levis2025galaxy,angthopo2019exploring,
salim2015green}. However, despite its importance, the observational 
definition of the GV remains ambiguous, and different selection methods often 
identify partially distinct galaxy populations.

A key source of this ambiguity is that different galaxy observables trace star 
formation over different timescales and therefore, respond differently to 
changes in star formation activity. The evolution of low-redshift galaxies 
moving from the blue cloud, through the GV, and onto the red sequence 
has been studied using data from the Galaxy And Mass Assembly Survey (GAMAS) 
over the redshift range $0.1 < z < 0.2$ \cite{driver2011galaxy,
phillipps2019galaxy}. Galaxies were classified according to their intrinsic
$(u^{*}-r^{*})$ colours derived from multi-wavelength spectral energy 
distribution fitting, which also provided estimates of stellar population 
ages and star formation timescales. The results suggest that most GV 
galaxies are consistent with a gradual decline in star formation 
over $\sim 2-4$~Gyr, likely driven by gas depletion rather than rapid 
quenching. Extrapolations of their star formation histories indicate 
continued reddening toward the passive red sequence, with no strong evidence 
for a recent change in their star formation time-scales, although internal or 
environmental processes may accelerate the decline in some cases. Ultraviolet 
(UV) emission traces recent star formation and rapid declination, typically
within $\sim 100$ Myr following quenching, whereas optical colours evolve 
more gradually and may remain blue long after the star formation rate (SFR) 
has begun to decrease. Consequently, UV-optical colour diagnostics are 
particularly sensitive to recently quenched or rapidly quenching systems, 
which may still occupy the blue cloud in purely optical diagrams
\cite{williams2009uvj, muzzin2013uvj}. In contrast, selections based solely 
on optical colours are susceptible to dust attenuation, which can redden 
actively star-forming galaxies and cause them to be misclassified as 
transitional systems \cite{salim2015green,battisti2016dust}. These effects 
illustrate the limitations of any single diagnostic and motivate the need for 
systematic comparisons across multiple observational tracers.

The GV population is further influenced by morphology and environment, 
which increases the complexity \cite{smith2022galaxy,bremer2018galaxy,
privatus2025ageing,kelvin2018galaxy,smethurst2015galaxy,jian2020redshift,
coenda2019green}. By examining how environment and redshift affect the 
evolution of the GV population, Ref.~\cite{jian2020redshift} found that the 
GV fraction generally remains below $20\%$ across all environments and 
redshifts, with higher values observed in the field than in groups and 
clusters. This fraction declines with decreasing redshift and increasing 
stellar mass. When focusing only on non-quiescent galaxies, the effective 
GV fraction is found to be higher in dense environments, suggesting more 
efficient quenching in groups and clusters. Additionally, galaxies in dense 
environments show lower specific star formation rate (sSFR) by 
$\sim 0.1-0.3$~dex. The findings of this study further highlight that 
quenching driven by a dense environment is active from $z \sim 1$.

Using the data from the GAMAS with stellar mass in the range 
$10.25 < \log(M_\star/M_\odot) < 10.75$ and redshift at $z < 0.2$, 
Ref.~\cite{bremer2018galaxy} carried out a study to investigate the influence 
of morphological evolution on galaxy quenching. The results of this study show 
that most GV galaxies contain both bulge and disc components, with the 
transition from blue cloud to red sequence mainly driven by changes in the 
disc rather than the bulge. These findings support a scenario in which 
galaxies evolve through 
the GV via gradual disc fading, with a substantial bulge already in place 
before the decline in star formation. The inferred GV crossing timescale is 
$\sim 1-2$~Gyr and shows little dependence on the environment, suggesting that 
gradual gas depletion primarily drives the evolution toward passive systems.

Observational studies show that many GV galaxies display intermediate 
morphologies, linking disc-dominated and bulge-dominated systems 
\cite{coenda2019green,smith2022galaxy}. Correlations between quenching 
efficiency and structural properties such as stellar mass concentration, 
bulge prominence, and central stellar density suggest that morphological 
transformation may be closely linked to the suppression of star formation 
\cite{smith2022galaxy}. Ref.~\cite{smith2022galaxy} examined the morphological 
properties of GV galaxies using data from the GAMAS combined with the Galaxy 
Zoo citizen-science project data \cite{lintott2008galaxy,willett2013galaxy}.
The study of Ref.~\cite{smith2022galaxy} also compared the 
structures of GV galaxies with those of blue star-forming and red quiescent 
populations to investigate the physical mechanisms driving their evolution. 
The results from this study also highlights that the spiral arm structure 
becomes weak as the galaxy transitions from blue cloud to red sequence. 
Furthermore, they observed that more ring structures are found for GV 
galaxies than in the blue cloud and the red sequence. These findings highlight 
the evidence of morphological evolution with quenching of star formation. 

Using the recently improved survey data sets and statistical methods, 
new avenues have opened for studying the GV galaxies. This includes the use 
of machine-learning approaches, involving neural networks, which have been used 
to classify galaxy morphologies \cite{angthopo2024retrieval,siudek2018vimos,
bluck2022quenching,ghosh2020galaxy,aguilar2025morphological,
sanjaripour2025selection}. Unsupervised clustering techniques have also been 
employed to isolate GV-like populations directly within multi-dimensional 
parameter space, eliminating the reliance on predefined colour-based 
selection criteria. While these techniques offer valuable new perspectives, 
they also highlight persistent inconsistencies between traditional GV
definitions \cite{schawinski2014green}. Ref.~\cite{turner2021synergies} 
investigated the origin and evolution of galaxy colour bimodality using 
unsupervised machine-learning techniques applied to observational data 
at two different cosmic epochs, i.e., the galaxy samples are analysed at 
$z\sim0.06$ and $z\sim0.65$. The low-redshift sample was taken from the 
GALEX-SDSS-Wide-field Infrared Survey Explorer (WISE) Legacy Catalogue, 
while the higher-redshift sample was obtained from the VIMOS Public 
Extragalactic Redshift Survey (VIPERS) \cite{guzzo2014vimos}. 
For both samples, galaxies were studied using nine independent 
rest-frame colour indices, spanning wavelengths from UV to the 
near-infrared (NIR). In both datasets, the galaxies were 
consistently divided into seven clusters, 
including four groups dominated by star-forming systems encompassing most GV 
galaxies and three groups composed mainly of passive galaxies. The clustering 
reveals that star-forming and GV galaxies form continuous morphological 
sequences at both redshifts, consistent with gradual, internally driven bulge 
growth that is associated with quenching at high stellar masses. 
Additional environmental effects are inferred only at low redshift, where they 
appear to influence the evolution of low-stellar mass passive galaxies.
In many cases, galaxies identified as GV members by clustering or 
machine-learning approaches do not align cleanly with conventionally defined 
colour-selected samples, suggesting that the GV may not correspond to a 
single evolutionary pathway but instead represents a heterogeneous collection 
of systems temporarily occupying similar regions of parameter space.

Although many studies have examined the GV, a thorough and quantitative 
comparison of widely used one-dimensional diagnostics within a consistent 
dataset remains lacking. In particular, it is still unclear how strongly 
different selection criteria overlap, whether they preferentially identify 
distinct subsets of galaxies, and how these selections are distributed across 
key observational planes such as the colour--magnitude, colour--stellar mass, 
and star formation--stellar mass relations. Resolving this issue is essential 
for assessing the reliability of GV definitions and for determining whether 
they genuinely trace galaxies in transitional phases of star formation or 
instead combine physically distinct systems that appear similar when described 
by one-dimensional indicators.

In this work, we perform a systematic comparison of different observational 
definitions of the GV, focusing on both their derived properties and their 
physical implications. We construct GV subsamples using several commonly 
adopted single-parameter diagnostics, including UV-optical colours, 
$D_n(4000)$, and sSFR thresholds. A key strength of this study is the use of 
galaxy's physical properties derived from the simultaneous modelling of UV 
and optical emission, providing a homogeneous and self-consistent set of 
stellar masses, SFR, and sSFR across the entire sample. 
This multi-wavelength approach offers a more robust basis for comparing 
different GV selection methods in comparison to analyses relying on a single 
wavelength regime. We compare these selections in terms of their stellar mass 
distributions, SFR, sSFR, and their locations within two-parameter diagnostic 
planes, including the colour-magnitude, colour-stellar mass, and star 
formation-stellar mass diagrams. By systematically examining each selection 
across multiple parameter spaces, we evaluate the degree of diagnostic overlap, 
identify potential selection biases, and assess how effectively each method 
isolates galaxies occupying intermediate regions of parameter space. This 
study provides new observational constraints on the interpretation of the GV 
and contributes towards developing more consistent and physically motivated 
definitions of transitional galaxy populations for future studies.

The structure of this paper is as follows. Section~\ref{secII} describes the 
data sets used and how the galaxy sample is constructed. 
Section~\ref{secIII} presents the results of the study and discusses their 
implications, and Section~\ref{secV} provides a summary and the main 
conclusions of the study. Throughout this work, we adopt a flat
$\Lambda$CDM cosmology with $H_0 = 70$~km~s$^{-1}$~Mpc$^{-1}$, 
$\Omega_m = 0.3$, and $\Omega_\Lambda = 0.7$. 

\section{Data and Methodology} \label{secII}
\subsection{Data Sources and Sample Construction}
This study combines photometric, spectroscopic, and derived physical 
properties of galaxies from the GALEX \cite{martin2005galaxy} and the SDSS 
\cite{york2000sdss} surveys to investigate the properties of GV galaxies. 
Specifically, the analysis is based on the Reference Catalogue of Spectral 
Energy Distributions (RCSED) \cite{chilingarian2017rcsed} and the 
GALEX-SDSS-WISE Legacy Catalog (GSWLC) \cite{salim2016galex}. Together, these 
datasets provide homogeneous multi-wavelength observations and galaxy physical 
parameters that are well-suited for studying galaxies transitioning from the 
blue cloud to the red sequence.

The RCSED catalogue was adopted as the primary dataset because it provides 
uniformly processed spectroscopic and photometric measurements for 
approximately $8\times10^{5}$ galaxies. RCSED combines UV, optical, and 
near-infrared observations from GALEX, SDSS, and the United Kingdom Infrared 
Telescope Deep Sky Survey (UKIDSS) \cite{lawrence2007ukirt}, 
providing homogeneous photometric and spectroscopic measurements, including 
the value-added properties such as emission-line, stellar population, 
morphological classifications, and spectroscopic redshifts. The accurate 
celestial coordinates and homogeneous data products available in RCSED make 
it an ideal reference catalogue for constructing the galaxy sample.

To obtain the galaxies' stellar masses, SFRs, and dust attenuations, the RCSED 
sample was cross-matched with the GSWLC, which combines UV photometry from the 
GALEX, optical photometry from SDSS, and infrared observations from the 
WISE \cite{wright2010wide}. The catalogue derives stellar masses, SFRs, and 
dust attenuation parameters through Bayesian spectral energy distribution 
(SED) fitting, simultaneously modelling the UV and optical emission while 
accounting for dust attenuation, metallicity, and different star formation 
histories. In this work, the SED-derived stellar masses and SFRs were adopted 
from GSWLC to characterise the stellar content and star-forming activity of 
galaxies.

The RCSED and GSWLC catalogues were cross-matched using their celestial 
coordinates (right ascension and declination) using a matching radius of $3$ 
arcseconds adopted to identify common sources, and the resulting matches 
were further verified using spectroscopic redshifts to minimise false 
associations. The matched catalogue was subsequently combined with the 
corresponding $D_{\mathrm{n}}(4000)$ measurements from the Max Planck 
Institute for Astrophysics and Johns Hopkins University (MPA-JHU) included in 
the SDSS DR17 \cite{abdurro2022seventeenth}. Only galaxies with reliable 
positional matches and the required photometric, spectroscopic, and derived 
physical parameters were retained for further analysis. The resulting sample 
is limited to galaxies within the redshift range $0.01 < z < 0.30$, with 
$14.15 \lesssim m_\mathrm{NUV} \lesssim 26.14~\mathrm{mag}$ 
and 11.75 $\lesssim m_r \lesssim 24.78~\mathrm{mag}$, where $m_\mathrm{NUV}$, 
$m_r$ denotes the extinction-corrected apparent magnitude in the GALEX near-UV 
band and SDSS $r$-band, respectively. Foreground galactic extinction was 
corrected using the extinction law of Ref.\ \cite{seaton1979interstellar} and 
the magnitudes were $k$-corrected using the methods described in Ref.\
\cite{chilingarian2010kcor}. These corrections ensure that derived colours 
and luminosities reflect intrinsic galaxy properties.

The final sample provides a homogeneous dataset containing UV, optical, and 
infrared photometry together with spectroscopic redshifts, stellar masses, 
SFR, and stellar age indicators. These parameters form the basis for 
investigating GV galaxies using both single-parameter diagnostics, such as 
$D_{\mathrm{n}}(4000)$ and sSFR, and two-parameter diagnostic diagrams 
involving UV-optical colours, stellar mass, and star formation activity.

\subsection{Single-Parameter Definitions of the Green Valley}\label{secIIB}
As already stated, the GV comprises galaxies with intermediate star-formation
activity, occupying the region between blue cloud and red sequence in
colour-magnitude and colour-stellar mass space. Because no single observable 
uniquely captures this transitional phase, we adopt four widely used 
one-dimensional diagnostics: rest-frame $\mathrm{NUV}-r$ colour, optical 
$u-r$ colour, the $D_n(4000)$ spectral index, and sSFR.

\begin{figure}[!h]
	\centering
	\includegraphics[width=0.45\linewidth]{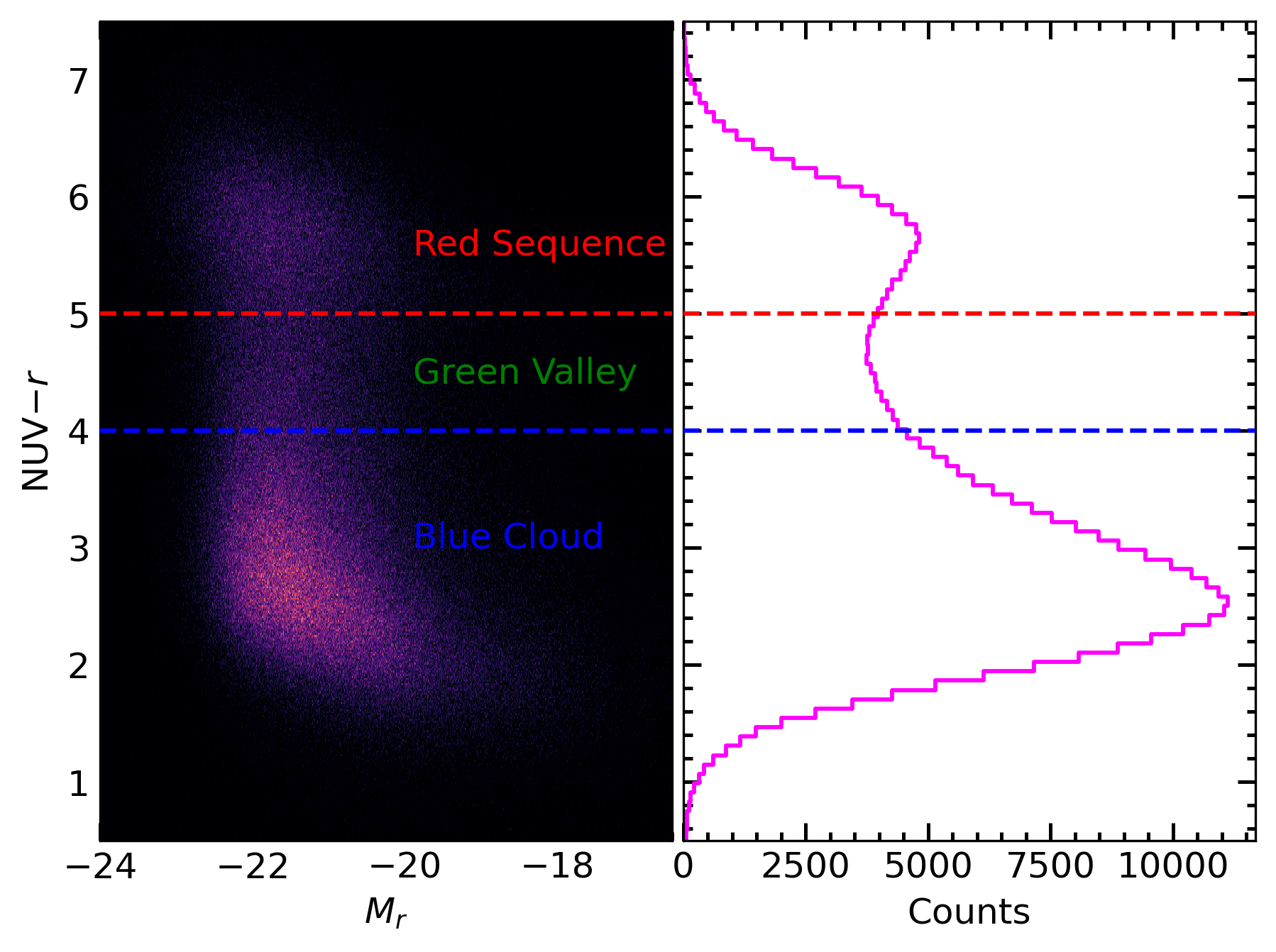}
	\includegraphics[width=0.46\linewidth]{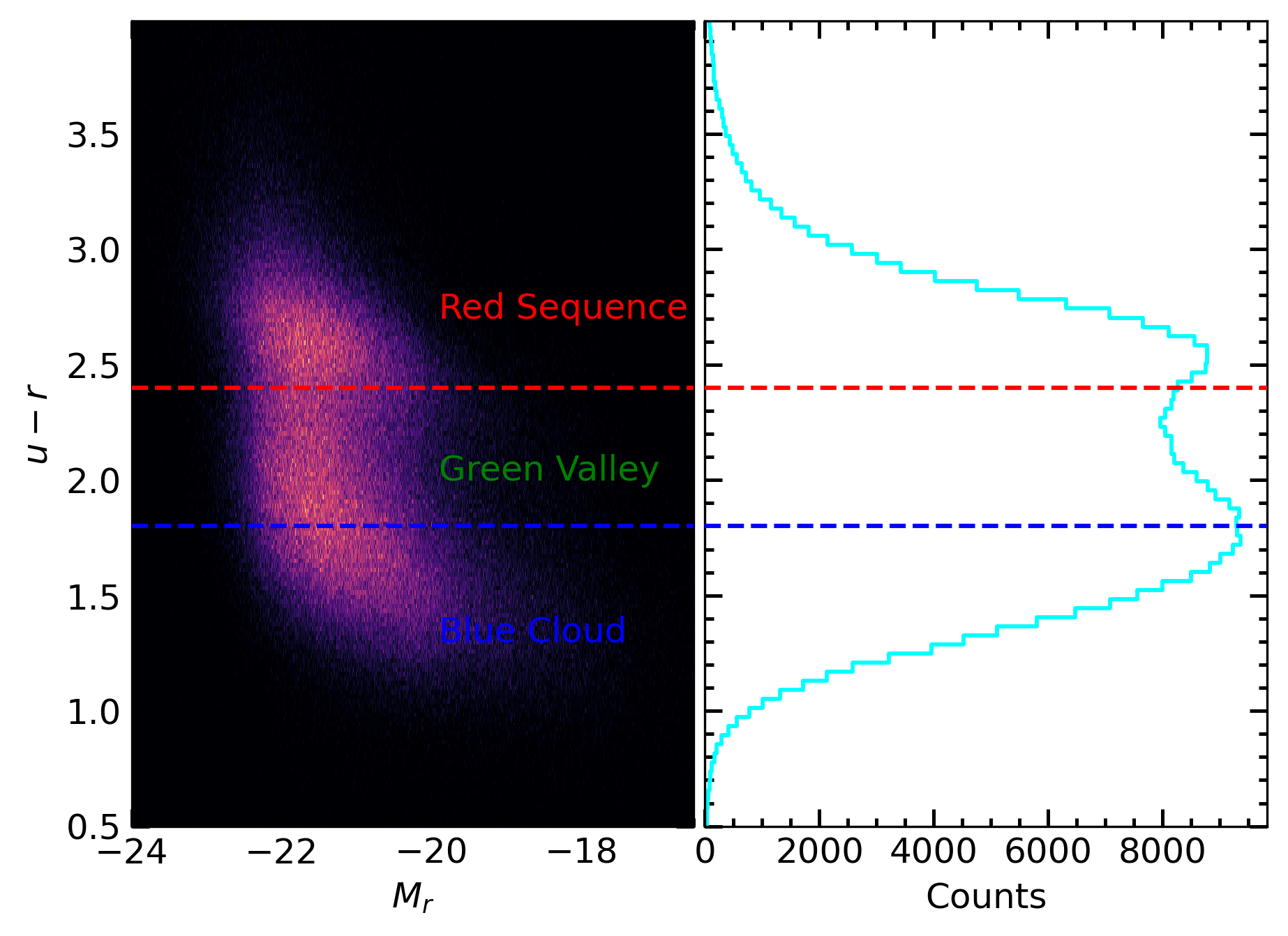}
        \vspace{-0.3cm}
	\caption{Illustration of $\text{NUV}-r$ (left panel) and $u-r$ (right 
panel) single-parameter diagnostics adopted to identify GV 
galaxies. The right-hand histogram for each panel in this figure and other 
similar figures shows the corresponding distributions, emphasizing the 
intermediate locus associated with transitional galaxies.}
	\label{gvd1}
\end{figure}

We first define a UV-optical GV sample using the rest-frame colour range
$4 < \mathrm{NUV}-r < 5$, corresponding to the intermediate region between 
the blue cloud and the red sequence as detailed in Ref.~\cite{salim2015green}, 
where a sample of $49706$ galaxies is obtained and shown in the left panel of 
Fig.~\ref{gvd1}. We also define an optically selected GV population using the 
colour range $1.8<u-r<2.4$, resulting in a sample of $129066$ galaxies as
shown by the right panel of Fig.~\ref{gvd1}.

\begin{figure}[!h]
	\centering
	\includegraphics[width=0.45\linewidth]{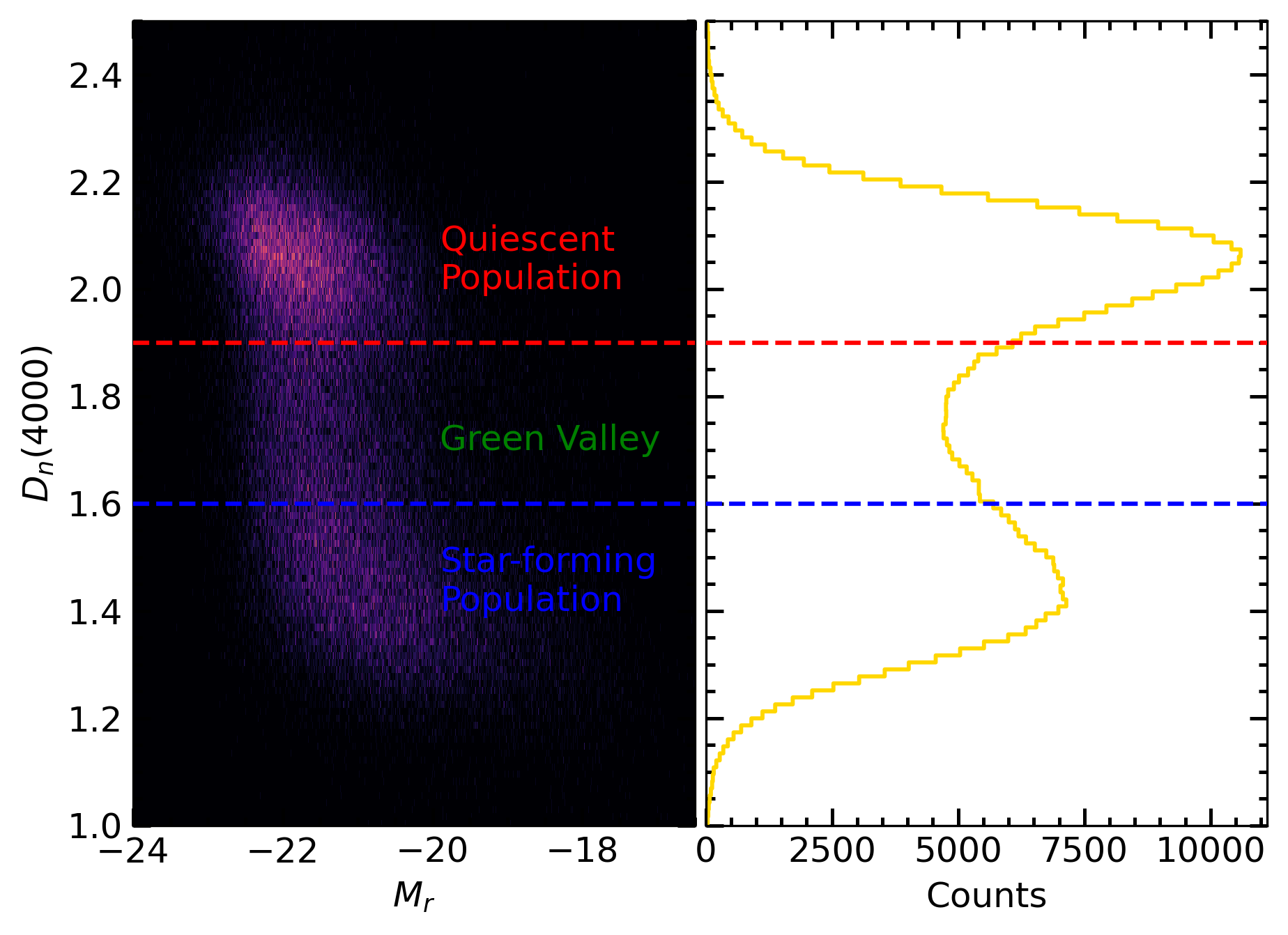}
	\includegraphics[width=0.47\linewidth]{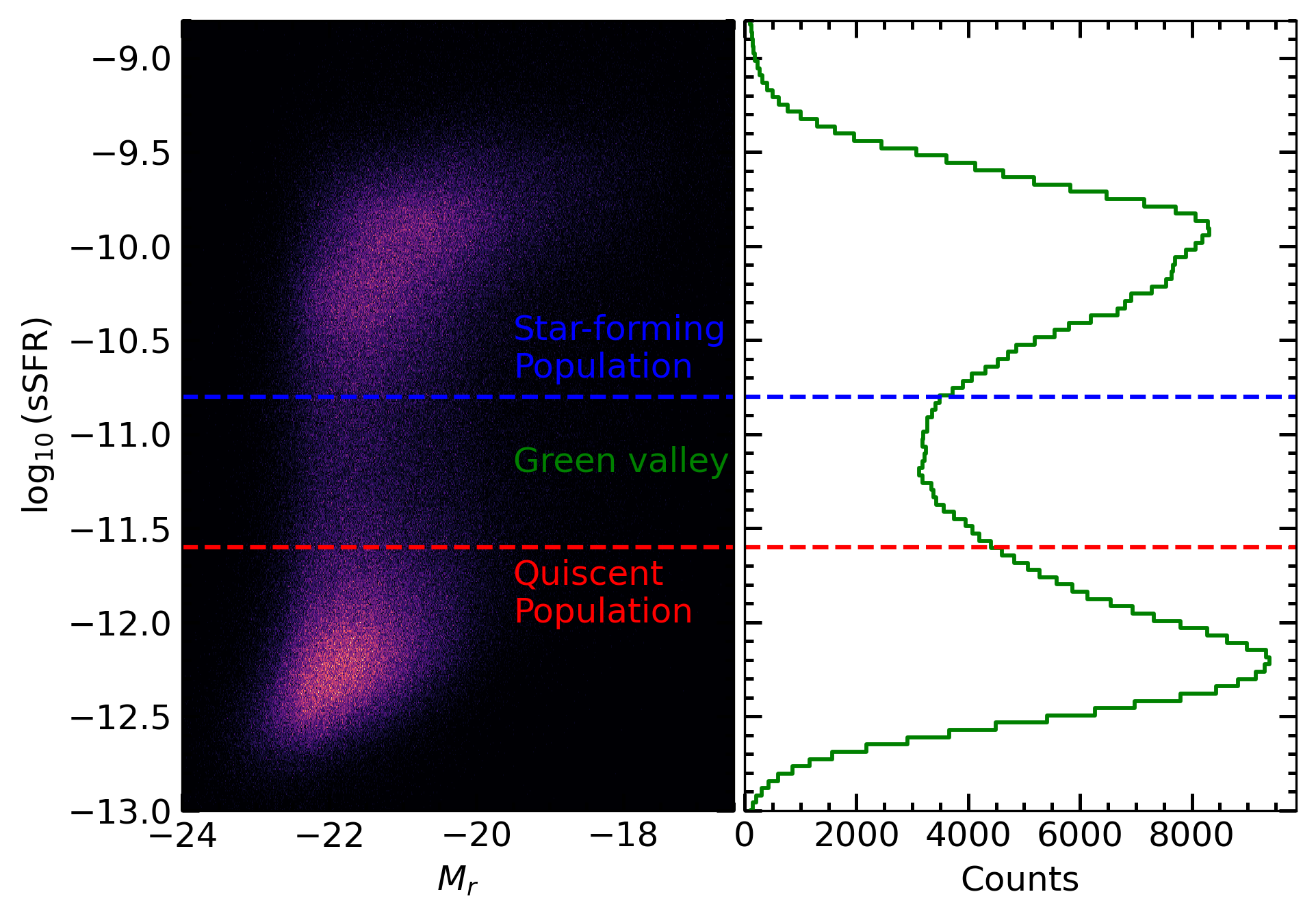}
        \vspace{-0.3cm}
	\caption{Illustration of the $Dn(4000)$ (left panel) and sSFR (right 
panel) diagnostics adopted to identify GV galaxies.}
	\label{gvd2}
\end{figure}

\begin{figure}[!h]
	\centering
	\includegraphics[scale=0.77]{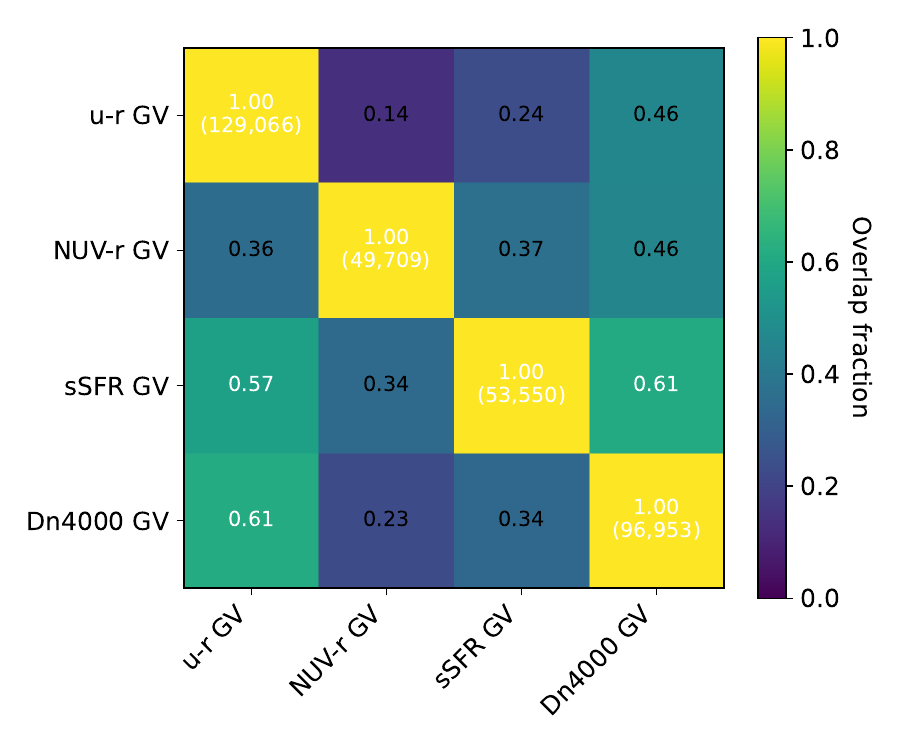}
        \vspace{-0.3cm}
\caption{Pairwise fractional overlap between GV samples selected 
using different single-parameter diagnostics: $u-r$ colour, $\mathrm{NUV}-r$
colour, sSFR, and the $D_n(4000)$ index.}
	\label{sgv}
\end{figure}

As an alternative to colour-based selections, we identify GV galaxies using 
the $D_n(4000)$ index, which traces the luminosity-weighted age of the 
stellar population. We define the GV as the galaxies having 
$1.5 < D_n(4000) < 1.8$ following the Ref.~\cite{kauffmann2003stellar}. This 
criterion yields $53550$ galaxies shown in the left panel of Fig.~\ref{gvd2}. 
Finally, we define a GV sample using sSFR, selecting galaxies with 
$-11.6 < \log_{10}(\mathrm{sSFR/yr^{-1}}) < -10.8$, consistent with previous 
studies \cite{salim2007uv,schawinski2014green}. This selection yields $96953$ 
galaxies shown in the right panel of Fig.~\ref{gvd2}. The agreement between 
the different single-parameter GV definitions is quantified using their 
pairwise fractional overlaps, as shown in Fig.~\ref{sgv}. Each element of the 
matrix represents the fraction of galaxies selected as GV by the criterion in 
the row that are also classified as GV by the criterion in the column. The 
diagonal elements represent the self-overlap of each criterion and therefore 
correspond to 100\%.


\section{Results and Discussion}\label{secIII}
\begin{figure}[!h]
	\centering
	\includegraphics[scale=0.58]{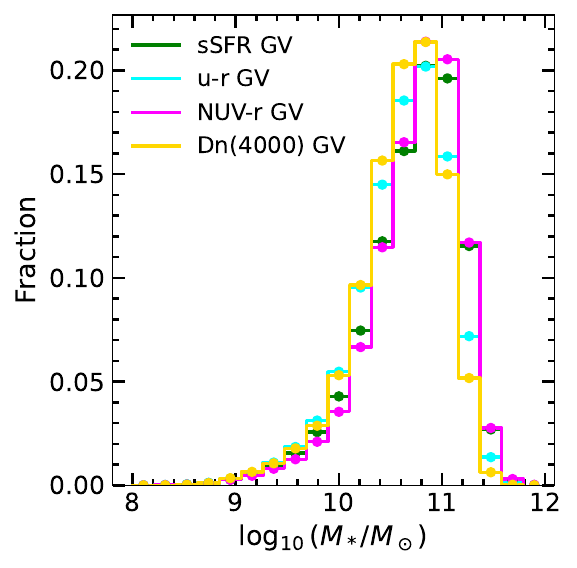}
	\includegraphics[scale=0.58]{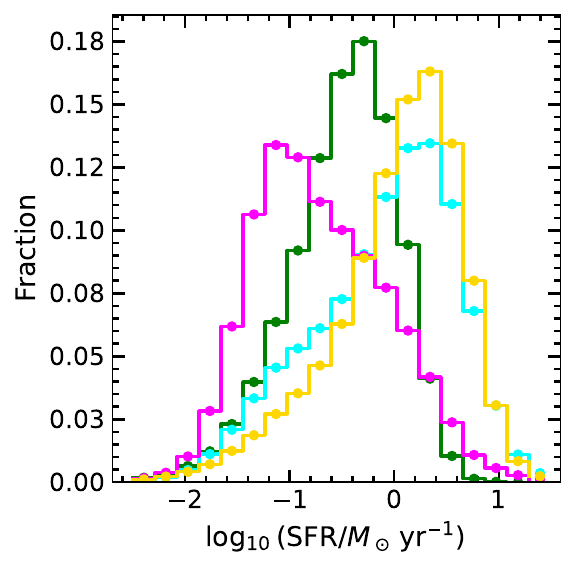}
	\includegraphics[scale=0.58]{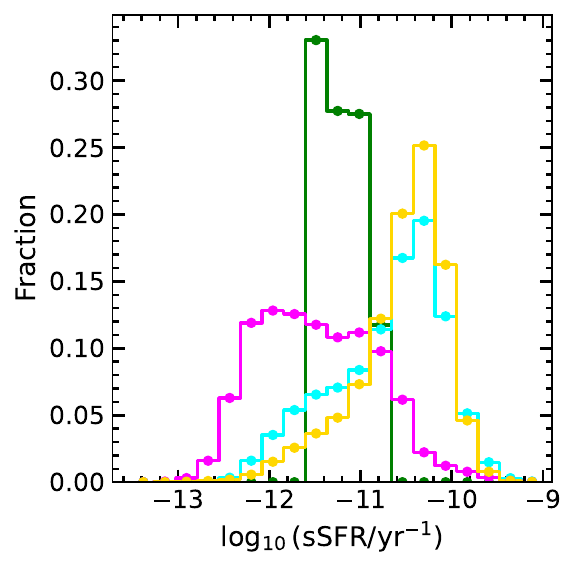}
        \vspace{-0.2cm}
	\caption{Distributions of $M_\star$ (left panel), SFR (middle panel), 
		and sSFR (right panel) for GV galaxies selected 
		using different diagnostics: sSFR (green), $u-r$ colour (cyan), 
		$\mathrm{NUV}-r$ colour (magenta), and $D_n (4000)$ (yellow). 
	}
	\label{dst}
\end{figure}

Figure~\ref{dst} shows the distributions of stellar mass ($M_\star$), SFR, 
and sSFR for galaxies identified as GV systems using four different criteria: 
$u-r$, NUV$-r$, sSFR, and $D_n(4000)$ defined in Subsection \ref{secIIB} 
above. The corresponding median values and associated spreads (16th -- 84th 
percentiles) are listed in Table~\ref{gv_stats}. We performed a pairwise 
Kolmogorov-Smirnov (KS) statistic comparison (as detailed in 
Refs.~\cite{hodges1958significance,harari2009kolmogorov}) of the $M_\star$, 
SFR, and sSFR distributions for the four GV selections shown in 
Table~\ref{GVC}. Because of the large sample sizes, all comparisons formally 
reject the null hypothesis of identical parent distributions ($p\ll0.001$), as 
in this case the KS statistic ($D$) is used to quantify the magnitude 
of the differences between the distributions. In the following, we 
therefore examine how these GV definitions map onto the two-dimensional 
diagnostic planes, allowing a direct comparison of their projected behaviours 
in colour--magnitude, colour--$M_\star$, and SFR--$M_\star$ spaces.

\begin{table}[h!]
	\centering
	\caption{$M_\star$, SFR, and sSFR median and their corresponding 16th 
and 84th percentiles for different GV definitions.}
	\label{gv_stats}
	\vspace{8pt}
	\setlength{\tabcolsep}{1.5pc} 
	\scalebox{1.0}{%
		\begin{tabular}{cccc}
			\toprule
			\toprule
			GV Definition & $\log_{10}M_\star(M_\odot)$& $\log_{10}\text{SFR}(M_\odot\,\mathrm{yr}^{-1})$ & $\log_{10}\mathrm{sSFR}(\mathrm{yr}^{-1})$ \\
			\midrule
			sSFR GV & $10.783^{+0.355}_{-0.517}$ & $-0.4358^{+0.441}_{-0.565}$ & $-11.227^{+0.295}_{-0.267}$ \\[3pt]
			$u-r$ GV & $10.679^{+0.369}_{-0.495}$ & $-0.0003^{+0.558}_{-0.665}$ & $-10.582^{+0.442}_{-0.889}$ \\[3pt]
			NUV$-r$ GV & $10.804^{+0.338}_{-0.474}$ & $-0.7867^{+0.767}_{-0.574}$ & $-11.536^{+0.742}_{-0.646}$ \\[3pt]
			$D_n(4000)$ GV & $10.657^{+0.344}_{-0.460}$ & $+0.1178^{+0.470}_{-0.752}$ & $-10.461^{+0.336}_{-0.630}$ \\
			\bottomrule
		\end{tabular}%
	}
\end{table}
\begin{table}[h!]
	\centering
	\caption{Pairwise Kolmogorov-Smirnov (KS) statistic comparison of the 
$M_\star$, SFR and sSFR distributions.}
	\label{GVC}
	\vspace{8pt}
	\setlength{\tabcolsep}{1.5pc} 
	\scalebox{1.0}{%
		\begin{tabular}{cccc}
			\toprule
			\toprule
			Comparison & $D(M_\star)$ & $D(\mathrm{SFR})$ & $D(\mathrm{sSFR})$ \\
			\midrule
			sSFR vs $u-r$          & 0.100 & 0.337 & 0.620 \\[2pt]
			sSFR vs NUV$-r$        & 0.030 & 0.242 & 0.467 \\[2pt]
			sSFR vs $D_n(4000)$    & 0.135 & 0.417 & 0.739 \\[2pt]
			$u-r$ vs NUV$-r$       & 0.121 & 0.381 & 0.459 \\[2pt]
			$u-r$ vs $D_n(4000)$   & 0.038 & 0.089 & 0.119 \\[2pt]
			NUV$-r$ vs $D_n(4000)$ & 0.153 & 0.469 & 0.578 \\
			\bottomrule
		\end{tabular}
	}
\end{table}

Here, we present the single-parameter GV galaxy samples defined already 
in the commonly used diagnostic planes, including colour--magnitude, 
colour--$M_\star$, and SFR--$M_\star$ relations.
For each GV definition, we present the selected galaxies onto these planes
and quantify their location relative to the blue cloud, the GV
region, and the red sequence. 
We begin by establishing the loci of the blue and red populations using 
the optical colour--magnitude relations of 
Refs.\ \cite{blanton2005properties,dhiwar2023witnessing,privatus2025mapping}. 
These sequences are described by
\begin{align}
	(g-r) &= 0.65 - 0.03(M_r + 20), \label{CM1} \\
	(g-r) &= 0.80 - 0.03(M_r + 20), \label{CM2}
\end{align}
where $M_r$ is the SDSS $r$-band absolute magnitude. Systems below 
Eq.~\eqref{CM1} correspond to actively star-forming galaxies, while those 
above Eq.~\eqref{CM2} trace the quiescent sequence.

The intermediate optical region is further better described in the 
colour--$M_\star$ plane, where the GV exhibits a $M_\star$-dependent slope. 
Thus, following the empirical calibration of 
Refs.\ \cite{schawinski2014green,privatus2025ageing,privatus2025main}, 
we adopt the relations,
\begin{align}
	u-r &= -0.24 + 0.25\log_{10}(M_\star/M_\odot), \label{gv1}\\
	u-r &= -0.75 + 0.25\log_{10}(M_\star/M_\odot), \label{gv2}
\end{align}
which provide the upper and lower boundaries of the GV in $u-r$ colour as a 
function of $M_\star$. These relations naturally track the gradual reddening 
of stellar populations with increasing $M_\star$.

Following previous studies of Refs.~\cite{noeske2007ms,elbaz2007ms,salim2014ms},
we quantify the offset from the main sequence (MS) of star formation. 
This classification scheme is motivated by an earlier work that separate
starburst (SB), MS, green valley, and quiescent systems 
\cite{rodighiero2011starburst}. For each galaxy, we measure its 
displacement from the MS by defining the residual
\begin{equation}
	\Delta \mathrm{SFR} = \log_{10}(\mathrm{SFR}_{\mathrm{gal}}) - \log_{10}(\mathrm{SFR}_{\mathrm{MS}}),
\end{equation}
where $\mathrm{SFR}_{\mathrm{gal}}$ is the observed SFR and 
$\mathrm{SFR}_{\mathrm{MS}}$ is the expected value of it for a MS galaxy 
of the same stellar mass. Galaxies with negative $\Delta \mathrm{SFR}$ values 
lie below the MS and exhibit progressively suppressed star formation, while 
systems with $\Delta \mathrm{SFR} \approx 0$ populate the MS. Positive offsets 
indicate enhanced star formation, typical of galaxies lying on the upper 
envelope of the MS or undergoing starburst activity. Based on their position 
in the $\mathrm{SFR}$--$M_\star$ plane, galaxies are classified using 
$\Delta \mathrm{SFR}$ as follows:
\begin{itemize}
	\item $\Delta \mathrm{SFR} > 0.5$ dex: SB galaxies, 
	undergoing elevated and likely short-lived episodes of intense 
	star formation (with typical timescales of $\sim 10^{8}$ yr).
	\item $-0.5 \leq \Delta \mathrm{SFR} \leq 0.5$ dex: MS star-forming 
galaxies, whose SFRs are close to the typical expected SFR for their stellar 
masses.
	\item $-1.1 \leq \Delta \mathrm{SFR} < -0.5$ dex: GV galaxies, 
	occupying the transitional regime between active star formation and 
        quiescence.
	\item $\Delta \mathrm{SFR} < -1.1$ dex: quiescent galaxies, 
	representing systems with strongly suppressed star formation.
\end{itemize}
We adopted the UV-based GV selection criteria in the NUV$-r$ versus stellar 
mass plane using a relation retrieved from Ref.\ \cite{coenda2019green}, 
given by
\begin{equation}
	(\mathrm{NUV}-r) = 0.92\,\log_{10}(M_\star/M_\odot) - 5.52 \pm 0.5,
	\label{coenda_relation}
\end{equation}
where the $\pm\, 0.5$~mag scatter defines a tilted GV band in the UV-stellar 
mass plane. In this relation, the slope reflects the increasing contribution of 
older stellar populations in more massive galaxies, providing a physically 
motivated boundary for separating transitional systems.

\begin{figure}[!h]
	\centering
	\includegraphics[scale=0.59]{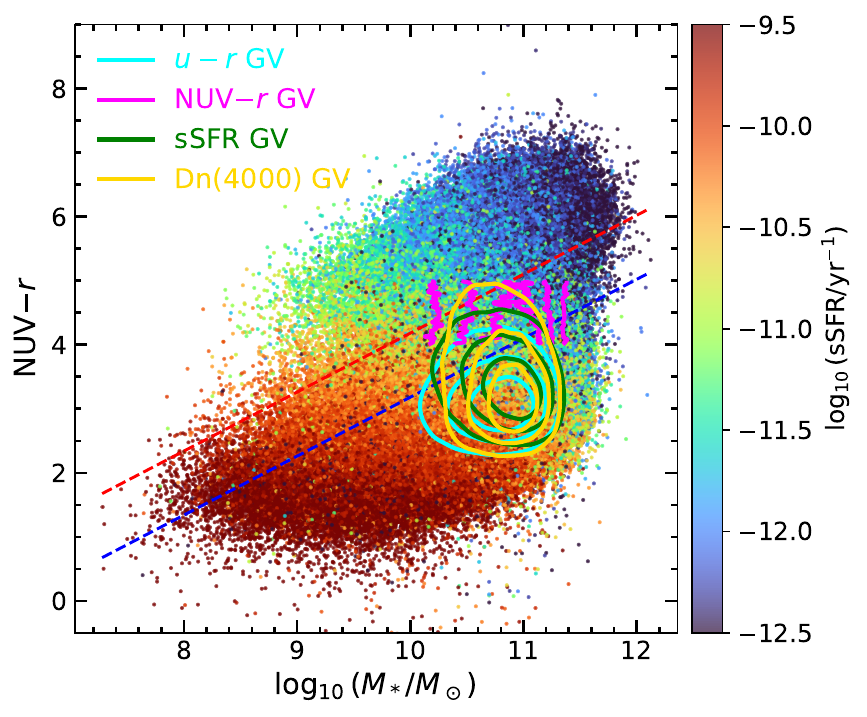}
	\includegraphics[scale=0.59]{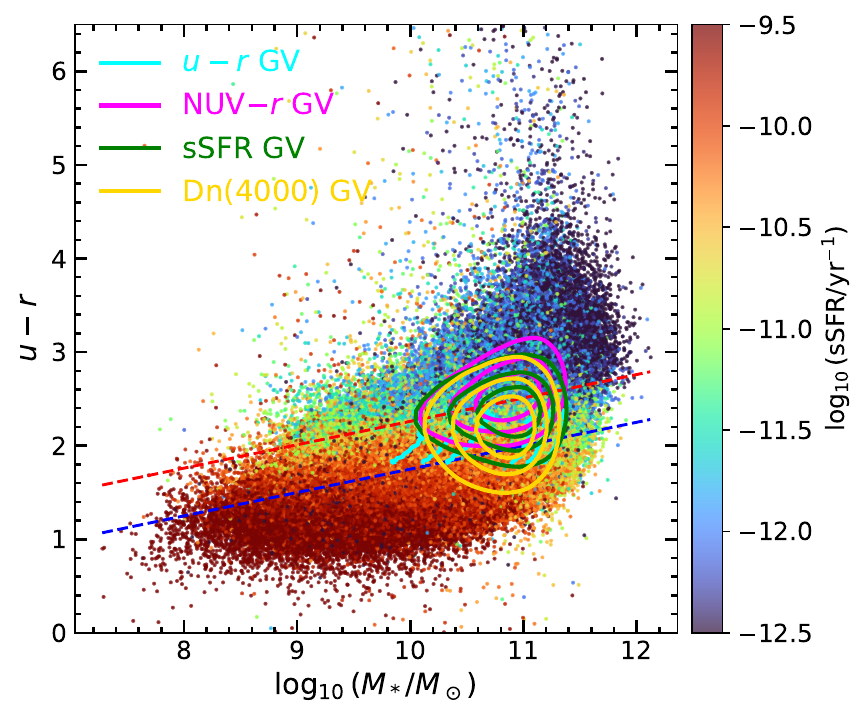}
        \vspace{-0.2cm}
\caption{Colour--stellar mass diagrams showing NUV$-r$ vs.\ stellar mass 
(left) and $u-r$ vs.\ stellar mass (right). Galaxies are colour-coded by
	$\log_{10}(\mathrm{sSFR/yr^{-1}})$. Overlaid contours show GV 
        selections based on $u-r$, NUV$-r$, sSFR, and $D_n(4000)$, 
        illustrating the level of overlap and divergence between the 
        different diagnostic definitions.}
\label{ALL_sSFR1}
\end{figure}

\begin{figure}[!h]
	\centering
	\includegraphics[scale=0.59]{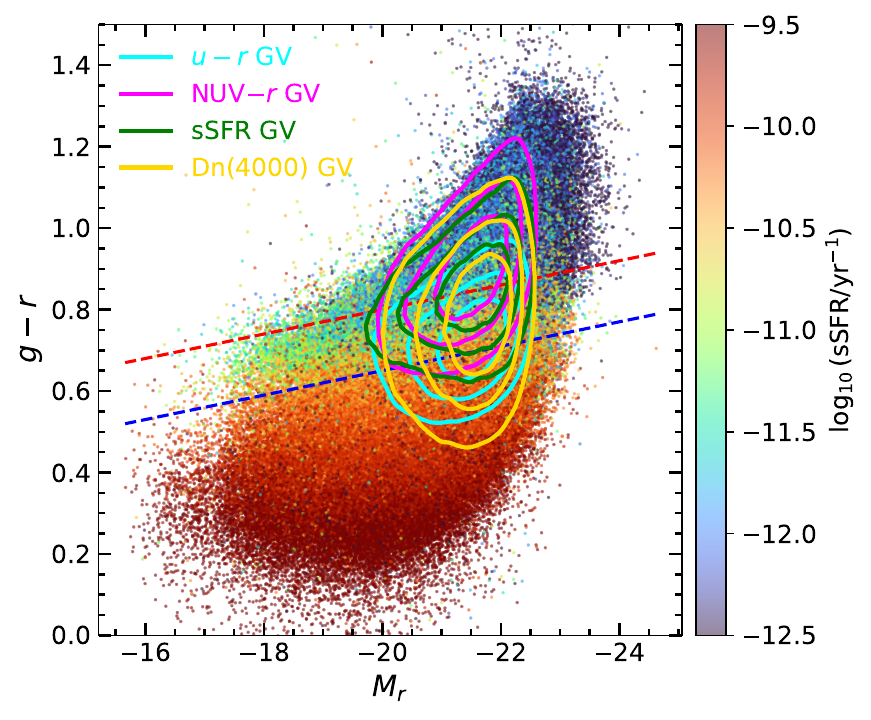}
	\includegraphics[scale=0.59]{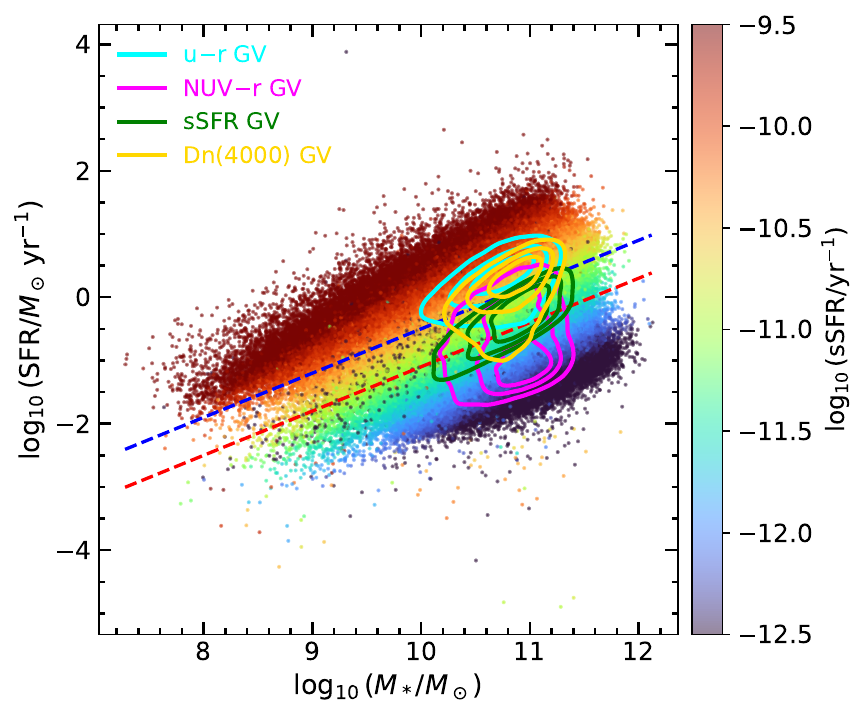}
        \vspace{-0.2cm}
	\caption{Left: colour--magnitude diagram ($g-r$ vs.\ $M_r$). 
		Right: SFR as a function of stellar mass. 
		In both panels, galaxies are colour-coded by 
                $\log_{10}(\mathrm{sSFR/yr^{-1}})$ 
		to illustrate how star formation activity is distributed 
		across these different parameter spaces.}
	\label{ALL_sSFR2}
\end{figure}

Together, the UV and optical relations outlined above provide the foundation 
for our two-parameter classification of GV galaxies. By presenting samples 
selected using a single parameter onto these two-dimensional parameter spaces, 
we assess their mutual consistency, identify potential biases and 
investigate whether different diagnostics select distinct subsets 
of transitioning galaxies. This multidimensional comparison underpins our 
interpretation of the GV population throughout this study. 
Figs.~\ref{ALL_sSFR1} and ~\ref{ALL_sSFR2} show the distribution of all 
single property GV-selected samples in the two-parameter diagrams, while the 
resulting statistics are summarized in Table~\ref{projection_statistics}.

\begin{table}
	\centering
	\vspace{8pt}
	\setlength{\tabcolsep}{6pt}
	\caption{Distribution of GV galaxies selected using 
different single-parameter diagnostics when projected onto colour--stellar 
mass, colour--magnitude, and star-formation planes. Galaxies are classified 
according to whether they lie above, within, and below the relevant GV region 
in each projection.}
	\label{projection_statistics}
		\vspace{7pt}
	\setlength{\tabcolsep}{1.2pc}
	\scalebox{1.0}{
		\begin{tabular}{lcccc}
			\toprule
			\toprule
			GV Definition &
			Total &
			Above GV &
			Within GV &
			Below GV \\
			\toprule
				\multicolumn{5}{c}{\textit{NUV$-r$ vs. $M_\star$}} \\
			\midrule
			$u-r$        & 129\,066 & 20\,882  (16.18\%) & 60\,332 (46.75\%) & 47\,852 (37.08\%) \\[2pt]
			NUV$-r$      & 49\,709 &  31\,077  (62.52\%) & 17\,724  (35.66\%) & 908  ( 1.83\%) \\[2pt]
			sSFR         & 53\,550 & 13\,853 (25.87\%) & 15\,784 (29.48\%) & 23\,913 (44.66\%) \\[2pt]
			$D_n(4000)$  & 96\,953 & 14\,656 (15.12\%) & 22\,235 (22.93\%) & 60\,062 (61.95\%) \vspace{0.3cm}\\
			\toprule
			
			\multicolumn{5}{c}{\textit{$u-r$ vs. $M_\star$}} \\
			\midrule
			$u-r$        & 129\,066 &  6\,726 (5.21\%) & 97\,795 (75.77\%) &  24\,545 (19.02\%) \\[2pt]
			NUV$-r$      & 49\,709 & 29\,296 (58.94\%) & 19\,754 (39.74\%)&  659 (1.33\%)\\[2pt]
			sSFR         & 53\,550 & 20\,636 (38.54\%) & 29\,768 (55.59\%) & 3\,146 (5.87\%) \\[2pt]
			$D_n(4000)$  & 96\,953 & 25\,492 (26.29\%) & 51\,530 (53.15\%) & 19\,931 (20.56\%) \vspace{0.3cm}\\
			\toprule
			
			\multicolumn{5}{c}{\textit{$g-r$ vs. $M_r$}} \\
			\midrule
			$u-r$        & 129\,066 & 13\,429 (10.40\%) & 23\,287 (18.04\%) &  92\,350 (71.55\%) \\[2pt]
			NUV$-r$      & 49\,709 & 10\,597 (21.32\%) & 34\,575 (69.55\%) & 4\,537 (9.13\%)\\[2pt]
			sSFR         & 53\,550 & 24\,485 (45.72\%) & 24\,040 (44.89\%) & 5\,025 (9.38\%) \\[2pt]
			$D_n(4000)$  & 96\,953 & 28\,211 (29.10\%) & 45\,845 (47.29\%) & 22\,897 (23.62\%) \vspace{0.3cm}\\
			\toprule
			
			\multicolumn{5}{c}{\textit{SFR vs. $M_\star$}} \\
			\midrule
			$u-r$        & 129\,066 & 79\,178 (61.35\%) &  27\,713 (21.47\%) & 22\,175 (17.18\%) \\[2pt]
			NUV$-r$      &49\,709 &  5\,937 (11.94\%)  & 13\,251 (26.66\%) & 30\,521 (61.40\%) \\[2pt]
			sSFR         & 53\,550 &  1\,225 (2.29\%)  & 33\,261 (62.11\%) & 19\,064 (35.60\%) \\[2pt]
			$D_n(4000)$  & 96\,953 & 44\,886 (46.30\%) & 27\,752 (28.62\%) & 24\,315 (25.08\%) \\[2pt]
			\bottomrule
		\end{tabular}
	}
\end{table}

The stellar mass distributions are similar across all four selection methods, 
with the median values around $\log_{10}(M_\star/M_\odot) \sim 10.66 - 10.80$, 
overall substantial overlap between their 16th and 84th percentile ranges and 
small pairwise KS statistics $D\sim 0.03 - 0.15$. These results suggest that 
the various single-parameter GV definitions largely probe galaxies within the 
same stellar mass regime, implying that stellar mass alone is not the 
primary driver of the observed differences between the samples.

For the star formation properties, a clear difference in distribution is 
observed wherein the NUV$-r$ selection method produces the galaxies with the 
lowest star formation activity having the median SFR, sSFR values of 
$\sim-0.79$, $\sim-11.54$, respectively. While $D_n(4000)$ selection produces 
the highest median SFR, sSFR ($\sim0.12$, $\sim-10.46$). The $u-r$ and sSFR 
selections lie between these two extremes; however, the $u-r$ sample shows 
slightly higher star formation activity with median values of $\sim-0.0003$, 
$\sim-10.52$ for the median SFR, sSFR, respectively, when compared to the sSFR 
defined GV sample with the median values of $\sim-0.44$, $\sim-11.23$ for SFR, 
sSFR, respectively. Correspondingly, the KS statistics of SFR and sSFR are 
larger when compared to the $M_\star$ attaining the maximum of 
$\sim 0.47$ ($u-r$ vs $D_n(4000$)), $\sim 0.74$ (sSFR vs $D_n(4000$)) for 
SFR and sSFR, respectively. Five of the six pairwise comparisons are 
classified as clearly different based on the KS effect size. This demonstrates 
that the different GV diagnostics preferentially select galaxies with 
distinct levels of ongoing star formation despite having similar 
$M_\star$ ranges. 

The width of the sSFR distributions further distinguishes the samples. The 
sSFR-selected sample shows a comparatively tighter distribution, as expected 
from a definition based directly on this parameter. Meanwhile, the 
colour-based ($u-r$, NUV$-r$) and spectral index ($D_n(4000)$) selections 
exhibit broader dispersions, reflecting the indirect connection between these 
observables and instantaneous star formation activity. These distinctions are 
also evident in the overlap fractions shown in Fig.~\ref{sgv}. The agreement 
between different GV definitions is generally modest. The strongest 
correspondence is found between the sSFR and $D_n(4000)$ selections, as well 
as between $u-r$ and $D_n(4000)$, with overlap fractions of $\sim 0.6$. In 
contrast, the NUV$-r$ selection shows relatively low overlap with the other 
methods (typically $\lesssim 0.4$), indicating that it isolates a more 
distinct subset of galaxies. These findings are the signs of the picture 
proposed by Ref.\ \cite{schawinski2014green}, in which it is argued that the 
GV is not a single homogeneous population but instead consists of galaxies 
following multiple quenching pathways. In particular, the wide separation 
observed in the sSFR distributions supports the interpretation that galaxies 
selected using different diagnostics represent different phases of star 
formation suppression rather than a unique transitional population. 
Similarly, Ref.\ \cite{salim2015green} demonstrated that UV colours are more 
directly linked to recent star formation than optical colours, making 
NUV$-r$ a more reliable tracer of quenching. Our results agree with this 
interpretation, as the NUV$-r$ selection is considerably more similar to the 
sSFR-selected sample in stellar mass than the optical $u-r$ selection, while 
simultaneously exhibiting systematically lower SFR and sSFR than the optical 
colour-selected galaxies. This suggests that NUV$-r$ is more effective at 
isolating galaxies undergoing genuine declines in star formation.

From the results shown in Table~\ref{projection_statistics} and
Figs.~\ref{ALL_sSFR1} -- \ref{ALL_sSFR2} supported by very small overall 
chi-square test p-value ($p<0.001$) when the number of galaxies above, 
within and below the green valley are compared, it reveals that the 
identification of GV galaxies is strongly dependent on the diagnostic used, 
with each selection tracing different aspects of galaxy evolution. This is 
clearly reflected in the redistribution of galaxies across colour--stellar 
mass, colour--magnitude, and SFR--stellar mass planes. In the NUV$-r$ versus 
stellar mass projection, the NUV$-r$-selected sample does not remain confined 
to the nominal GV region. Instead, a majority of galaxies ($\sim 62.5\%$) lie 
above the GV, while only $\sim 35.7\%$ fall within it. This pattern indicates 
that UV-based selection criteria extend significantly into the red sequence 
in this parameter space. In contrast, the $u-r$ selection appears more 
balanced, yielding comparable fractions of galaxies within and outside the 
green valley. The sSFR and $D_n(4000)$ selections are instead shifted toward 
lower NUV$-r$ values, with approximately $44.7\%$ and $62.0\%$ of their 
samples lying below the green valley, respectively. Overall, this highlights 
the strong dependence of UV colours on recent star formation activity, as well 
as their sensitivity to dust attenuation.

A different trend is observed in the $u-r$ versus stellar mass plane, 
where the $u-r$-selected sample is highly concentrated within the GV, with 
about $75.8\%$ of galaxies lying in this region. The sSFR and $D_n(4000)$ 
selections also exhibit substantial overlap with the GV, although both are 
more broadly distributed across the diagram. In contrast, the NUV$-r$ 
selection is dominated by galaxies located above the optical GV, with 
approximately $58.9\%$ of the sample falling in this region, indicating that 
galaxies with intermediate UV colours can still appear relatively red in 
optical bands. This discrepancy between UV and optical-based diagnostics 
highlights the joint impact of dust attenuation and recent star formation on 
observed galaxy colours. In the colour–magnitude diagram ($g-r$ versus $M_r$), 
the differences between the various diagnostics become more evident. The 
NUV$-r$ selection is most strongly concentrated within the optical GV, with 
approximately $69.6\%$ of galaxies falling in this region, whereas the $u-r$ 
selection is predominantly composed of galaxies below the GV, with about 
$71.6\%$ lying in the blue cloud, indicating significant contamination from 
star-forming systems. The sSFR-selected sample contains nearly equal fractions 
of galaxies in the green valley ($\sim44.9\%$) and the red sequence ($\sim45.7\%$), indicating the presence of an intermediate galaxy population. Meanwhile, the $D_n(4000)$-based selection covers a broad range, 
with substantial fractions both within and outside the GV, reflecting its 
sensitivity to a wide distribution of stellar population ages.

The SFR–stellar mass diagram provides the most physically straightforward 
representation of the galaxy population. When the GV is defined using sSFR, it 
produces a well-separated intermediate population, with roughly $62.1\%$ of 
galaxies located within the GV and only a small fraction ($\sim2.29\%$) 
lying above it. In contrast, the $u-r$ selection preferentially includes 
galaxies with higher SFRs, with about $61.4\%$ lying above the green valley. 
The NUV$-r$ selection shows the opposite trend, with a similar fraction 
($\sim 61.4\%$) falling below the GV, indicating a bias toward more passive 
systems. The $D_n(4000)$-based selection is instead widely distributed 
across the entire parameter space, encompassing both star-forming and quenched 
systems without a clear preference for either population.

These findings demonstrate that selecting GV galaxies using a single-parameter 
is highly dependent on the choice of diagnostic. The variations in the 
fractions reported in Table~\ref{projection_statistics} reflect the fact that 
each observable traces different physical timescales and underlying processes. 
Broadband optical colours are influenced by both dust attenuation and 
star-formation history, which can move galaxies between the blue cloud and 
red sequence depending on the projection \citep{Salim2007, Maller2009}. 
UV colours respond even more strongly to recent star formation and dust 
content, helping to explain the systematic offsets seen in the NUV$-r$ 
selection across different parameter spaces \citep{Martin2007}. 
The $D_n(4000)$ index is only weakly affected by dust and primarily reflects 
the luminosity-weighted age of the stellar population, evolving over 
relatively long timescales. Because of this, it picks out galaxies spanning a 
wide range of evolutionary stages, from actively star-forming systems to 
fully quiescent ones \cite{kauffmann2003stellar, brinchmann2004physical}. 
This behaviour leads to its broad spread across all projections. By contrast, 
the sSFR-based selection shows consistent behaviour across all parameter 
spaces. Since sSFR directly traces the current level of star formation, it 
picks out a well-defined intermediate population that remains identifiable 
regardless of the projection used. This is seen in its strong concentration 
within the GV in the SFR--stellar mass plane, along with a more even 
distribution in both the colour--stellar mass and colour--magnitude diagrams.

The comparison with Ref.\ \cite{angthopo2019exploring} further supports the 
conclusion that different observational diagnostics identify partially 
overlapping but physically distinct GV populations. This reference study 
showed that optical colours, UV colours, and spectral indices are sensitive 
to stellar populations over different timescales, leading to systematic 
differences in the inferred GV membership. Our statistical analysis reaches 
the same conclusion, with significant differences observed in both the star 
formation properties and the occupancy of galaxies within colour--magnitude, 
colour--stellar mass, and SFR--stellar mass diagrams. Our results also agree 
with the multidimensional analysis of Ref.\ \cite{turner2021synergies}, who 
showed that no single observable provides a complete description of GV 
galaxies and that combining multiple diagnostics yields a more 
physically meaningful classification. The significant differences found 
between all pairwise GV definitions, particularly in SFR and sSFR, reinforce 
this conclusion by demonstrating that each selection criterion preferentially 
samples galaxies at different stages of quenching. Consequently, studies 
relying on a single GV definition should account for the systematic biases 
introduced by the adopted diagnostic when interpreting galaxy evolution.

\section{Summary and Conclusions}\label{secV}
In this study, we investigate how four commonly used one-dimensional 
definitions of the GV relate to one another when analysed across different 
parameter spaces and when projected to two-dimensional diagrams using galaxy
physical properties derived from the simultaneous modelling of UV and optical 
emission. We define GV samples using optical and UV colours, sSFR, and the 
$D_n(4000)$ index, and study their distributions within colour--stellar mass, 
colour--magnitude, and SFR--stellar mass planes. Our key findings are 
summarised below:
\begin{itemize}
	\item GV samples selected using different diagnostics exhibit
        substantially different projected distributions, demonstrating that 
        commonly used single-parameter cuts do not identify equivalent galaxy 
        populations. This is further supported by the generally modest overlap 
        between different definitions, indicating that each diagnostic isolates
        a partially distinct subset of galaxies.
	
	\item Colour-based selections show strong projection dependence.
	The NUV$-r$ definition forms a tighter selection population in 
        $g-r$ colour stellar mass, but shifts predominantly toward the red 
        sequence in $u-r$ colour stellar mass projections and toward lower 
        star-formation activity in the SFR--stellar mass plane, indicating a 
        bias toward more quiescent systems. In contrast, the $u-r$-based 
        sample provides a tighter selection in optical colour--stellar mass 
        space but shows a systematic shift toward higher SFRs, reflecting its 
        sensitivity to galaxies that retain significant star formation despite 
        intermediate optical colours.
	
       \item The $D_n(4000)$-based definition produces the most heterogeneous 
       sample. Although it is relatively insensitive to dust, it primarily 
       traces the age of the stellar population and therefore includes 
       galaxies covering a wide range of star-formation activity, from 
       actively star-forming to fully quiescent systems. 
	
      \item The sSFR-based selection behaves in a more uniform way across all 
       parameter spaces, outlining a clear intermediate sequence in the 
       SFR--stellar mass plane and showing a relatively even spread in both 
       colour--magnitude and colour--stellar mass diagrams. This suggests 
       that it is effective at picking out galaxies in the process of 
       transitioning in their star formation activity.
   
       \item Despite differences in their construction, the various 
        selections cover a similar range in stellar mass, indicating 
        that the observed variations are driven primarily by 
        star-formation properties rather than by stellar mass.
	
\end{itemize}
Taken together, these results demonstrate that GV identification based on a 
single observable is intrinsically diagnostic-dependent. Although the stellar 
mass distributions remain remarkably similar across all GV definitions, 
statistically significant differences in star formation properties and 
occupancy of diagnostic diagrams demonstrate that these definitions are not 
interchangeable. Each criterion samples galaxies at different evolutionary 
stages, implying that the choice of GV definition can introduce systematic 
biases into studies of galaxy quenching.

This study provides a statistically robust comparison of commonly adopted GV 
selection criteria using galaxy properties derived from the simultaneous 
modelling of UV and optical emission, ensuring a homogeneous and internally 
consistent analysis across all definitions. However, a complete understanding 
of the physical origin of the observed differences requires a multidimensional 
investigation. As an extension of this work, we will investigate the 
morphological properties, AGN incidence, environmental dependence, stellar 
metallicity, and cold gas content of the different GV populations. These 
quantities probe complementary aspects of galaxy evolution, including 
structural transformation, feedback processes, external environmental effects, 
chemical evolution, and gas depletion. Their joint analysis will enable us to 
determine whether the various GV definitions preferentially identify galaxies 
experiencing distinct quenching pathways or simply represent different 
observational manifestations of the same underlying evolutionary sequence. 
Such an analysis will provide stronger physical constraints on the mechanisms 
responsible for the transition of galaxies from the star-forming MS to the 
quiescent population.

\section*{Acknowledgements} PP acknowledges support from The Government of 
Tanzania through the India Embassy, Mbeya University of Science and Technology 
(MUST) for Funding. UDG is thankful to the Inter-University Centre for 
Astronomy and Astrophysics (IUCAA), Pune, India for the Visiting 
Associateship of the institute. This work is based on observations made with 
the Galaxy Evolution Explorer (GALEX), a NASA Small Explorer mission. 
GALEX is operated for NASA by
the California Institute of Technology under NASA contract NAS5-98034.
This research also makes use of data from the Sloan Digital Sky Survey
(SDSS). Funding for SDSS-III has been provided by the Alfred P. Sloan 
Foundation, the Participating Institutions, the National Science Foundation, 
and the U.S. Department of Energy Office of Science. SDSS-III is 
managed by the Astrophysical Research Consortium for the
Participating Institutions of the SDSS-III Collaboration including the
University of Arizona, the Brazilian Participation Group, Brookhaven
National Laboratory, Carnegie Mellon University, University of Florida,
the French Participation Group, the German Participation Group, Harvard
University, the Instituto de Astrofísica de Canarias, the Michigan State /
Notre Dame / JINA Participation Group, Johns Hopkins University, Lawrence
Berkeley National Laboratory, Max Planck Institute for Astrophysics,
Max Planck Institute for Extraterrestrial Physics, New Mexico State
University, New York University, Ohio State University, Pennsylvania
State University, University of Portsmouth, Princeton University,
the Spanish Participation Group, University of Tokyo, University of Utah,
Vanderbilt University, University of Virginia, University of Washington,
and Yale University.

\end{document}